\documentstyle[11pt,newpasp,twoside,psfig]{article}
\begin{document}

\title{Radio galaxies and the star formation history of the universe}
\author{Gopal-Krishna}
\affil{National Centre for Radio Astrophysics/TIFR, Post Bag 3, 
Ganeshkhind, Pune 411 007, India (krishna@ncra.tifr.res.in)}
\author{Paul J.\ Wiita and M.\ Angela Osterman}
\affil{Department of Physics \& Astronomy,  Georgia State University,
Atlanta GA, 30303-3088, USA (wiita,osterman@chara.gsu.edu)}

\setcounter{page}{111}
\index{Gopal-Krishna}
\index{Wiita, P. J.}
\index{Osterman, M. A.}

\begin{abstract}
Multi-wavelength observations made in the last decade 
suggest that the universe underwent an intense phase of star formation
in the past ($z > 1$).  This intensive activity is commonly attributed
to a higher galaxy merger rate when the universe was a fraction of its
present age.  We examine the role of the powerful radio sources whose
comoving density is known to be a few orders of magnitude higher
at $z \sim 2$, the `quasar era'.  Taking into account recent models 
for the temporal evolution of the size and luminosity of a powerful
double radio source, as well as $\Lambda$CDM simulations of the cosmic
web of baryonic material at different redshifts, we argue that during
the quasar era a high fraction of the volume of the web was occupied
by the lobes of double radio sources.  Widespread compression of
protostellar clouds, triggered by the high pressure of the synchrotron 
plasma of the radio lobes, can thus be expected to have played a
significant role in the global star formation history of the universe.  These
lobes can also yield a rather high level of magnetization of the intergalactic
medium at these early cosmic epochs.

\end{abstract}

\section{Introduction}
Flux density-limited samples of radio sources show that the comoving space
density of powerful radio galaxies (RGs) declined between 100 and 1000
times from redshifts of 2--3 to the present epoch (e.g.\ Willott et al.\
2001).  The star formation rate also peaked in roughly the same epoch
(e.g.\ Archibald et al.\ 2001).  Because of severe
adiabatic and inverse Compton losses, most
old and large radio galaxies are very difficult to detect in surveys
and only the youngest can be seen at high redshifts
(e.g.\ Blundell et al.\ 1999, hereafter BRW).
Cosmological simulations have indicated that most of the matter that 
will form galaxies by the current epoch was in the form of filaments
that filled only a small portion of the universe at those redshifts
(e.g.\ Cen \& Ostriker 1999).  Together, these facts lead us to conclude that 
the formation of many of those  galaxies may have been triggered by overpressured
radio lobes, which probably filled a substantial portion of those filaments then. 
 A preliminary discussion of this
work is in Gopal-Krishna \& Wiita (2001, hereafter GKW) and more extensive calculations
are underway (Osterman, Wiita, Gopal-Krishna \& Kulkarni, in preparation).

\section{Radio galaxy visibility}
All modern models of RG evolution (Kaiser et al.\ 1997;
BRW; Manolakou \& Kirk 2002) agree that radio
flux declines dramatically with increasing source size (adiabatic
losses) and with $z$ (inverse Compton losses off the cosmic background
radiation).  Theoretical distributions of RG powers, sizes, 
redshifts and spectral indices can be nicely matched by
models that require most RGs to remain active for $T \simeq 5 \times 10^8$y and to have
a distribution of jet powers ($Q_0$) that goes as $\sim Q_0^{-2.6}$
(BRW).  X-ray observations indicate  that the density of
the matter through which the jets propagate declines with distance
roughly as
$n(r) = n_0(r/a_0)^{-\beta}$, with  $n_0 = 1.0 
\times 10^4{\rm m}^{-3}$, $a_0 = 10~$kpc, and
$\beta = 1.5$.  This leads to the  linear size of the RG being given by
\begin{equation}
D(t) = 3.6 a_0 \Bigl(\frac{t^3 Q_0}{a_0^5 m_p n_0}\Bigr)^{1/(5-\beta)}.
\end{equation}

Using these models, we find that in most RGs, particularly those
at $z > 2$, the central engines remain active for much longer times than 
those galaxies are detected in flux limited surveys, and therefore they 
should grow to very large linear sizes (typically $D(T) > 1$ Mpc),
although detecting them would require extremely sensitive radio surveys
with redshifts.
From BRW we find that the visibility time, $\tau \propto Q_0^{1/2}$, 
and to properly estimate the actual number of RGs
from those detectable in flux-limited surveys, one must multiply
by a correction factor ($T/\tau$) of roughly 50 for powerful
RGs during the quasar era (GKW).

\section{Radio luminosity function -- RLF}
Most RLF studies are plagued by uncertainties resulting from incomplete
knowledge of the redshifts
of the radio sources, but results based 
upon the 3CRR, 6CE and 7CRS surveys of different flux limits 
have the advantage of having 96\% of their redshifts known (Willott
et al.\ 2001).  In addition, their selection at low frequencies minimizes
the bias due to  relativistic beaming.

The powerful  FR II sources are nearly
3 dex above the local RLF by $z \sim 2$, and their RLF varies little
out to the beginning of the quasar era at $z \sim 3$, 
while it appears to decline at higher $z$ 
(Willott et al.\ 2001).  Furthermore, the RLF for those redshifts is nearly
flat for over a decade in radio power above $P_{151} \ge 10^{25.5}$
W Hz$^{-1}$ sr$^{-1}$, which is where the FR II sources are most
numerous.  

Combining these results with the correction factor discussed in \S 2
we find that at $z = 2.5$ the actual proper density  of powerful 
radio sources born in an interval
$T$ is
$\rho \simeq 4 \times 10^{-5}
(1+z)^3 T_5~{\rm Mpc}^{-3} (\Delta \log P_{151})^{-1}$,
where $T_5 \equiv T/(5\times 10^8{\rm y})$.  We then integrate
 over the
roughly 1.25 dex of the peak of the RLF.  Finally, we take into
account the fact that several generations of RGs will be born 
and will die within the $\sim 2$ Gyr duration of the quasar era.  This leads
us to the total proper density, $\Phi$, of intrinsically powerful 
radio sources: $\Phi = 7.7 \times 10^{-3} ~{\rm Mpc}^{-3}$, which is
independent of the assumed value of $T$.

\section{The relevant universe}
Numerical models of the evolution of  $\Lambda$CDM universes
indicate that at $z \sim 0$, roughly 70\% of baryons are in a cosmic web of
filaments of warm-hot gas and embedded galaxies and clusters that together
occupy only about 10\% of the volume of the universe  (e.g.\ Cen \& Ostriker 1999;
 Dav{\'e} et al.\ 2001).  But at $z \sim 2.5$ the growing network of filaments
comprised only about 20\% of the baryonic mass, and a quite small fraction, $\eta \simeq 0.03$,
 of the total volume.

The massive galaxies that harbor supermassive black holes large enough to
form RGs at early times would have typically formed in the densest portions
of those filaments. The lobes ejected from them would mostly remain within
the filaments, and since it is in this relatively small, `relevant universe',
that new galaxies formed out of denser gas clumps, we only need to be concerned with what fraction of
this relevant universe the lobes permeated.  We find that the mean volume of
a radio source is $\langle V(T) \rangle \simeq 2.1 T_5^{18/7} ~{\rm Mpc}^{3}$, and 
thus,  the volume 
fraction of the  relevant universe which radio lobes born during
the quasar era cumulatively swept through is:
\begin{equation}
\zeta = \Phi ~\langle V (5 \times 10^8 {\rm yr}) 
\rangle ~(0.03/\eta)(5/R_T)^2 \simeq 0.5,
\end{equation}
where $R_T \sim 5$ is the typical width to length ratio of an RG.  The energy
density injected by the lobes into the filaments is $u \simeq 2 \times 10^{-16}$
J m$^{-3}$ for those same canonical parameters.  Because $\langle V(T) \rangle$ is
a sensitive function of $T$, if the typical RG lifetime is  $< 10^8$ yr then
$\zeta < 0.01$ and $u < 9 \times 10^{-18}$ J m$^{-3}$(GKW).

\section{Radio lobe triggered  star formation}
The discovery of the alignment effect between extended optical emission lines and
radio lobe directions (e.g.\ McCarthy et al.\ 1987; Chambers et al.\ 1987) quickly led to
calculations (e.g.\ Begelman \& Cioffi 1989; Rees 1989) that indicated that
star formation could be triggered by these expanding overpressured lobes.
Recent  hydrodynamical simulations including cooling (Mellema et al.\ 2002) confirm 
that this is likely to occur through cloud fragmentation, cooling and compression.
{\sl HST} observations of high-$z$ RGs and associated
optical emission (e.g.\ Best et al.\ 1996) 
support this scenario.

We further estimate (GKW) that these  powerful RGs create lobes which 
typically remain rapidly expanding, with overpressures of factors
exceeding 100 (or Mach numbers above 10) out to distances
of well over 1 Mpc.  Supersonic expansion into a two-phase circumgalactic 
medium will compress many of the cooler gas clumps, rapidly reducing
the Jeans mass by factors of 10--100 and thereby 
triggering starbursts (Rees 1989; Mellema et al.\ 2002).

\section{Conclusions} 
Although the  local universe is very sparsely populated by powerful radio sources, 
several large factors work in the same direction
 to make them remarkably
important for galaxy formation during the quasar era.   
First, their comoving density was roughly
1000 times higher at $\sim 2 < z < \sim 3$.  Second, only a small fraction (roughly
two percent) of the powerful sources present during that period are detected
in the surveys used to produce the RLFs, because of severe inverse 
Compton and adiabatic losses; these unseen radio lobes
fill very large volumes.  Third, the fraction
of the volume of the universe occupied by the material
during the quasar epoch that would finally condense into clusters of
galaxies was only a few percent, so these lobes only had to permeate this `relevant
universe' rather than the entire universe.  In that the best models of RG evolution
indicate that the lobes  are overpressured and supersonically expanding into
the relevant universe, the scenario that many massive starbursts, and even 
many galaxies, are formed in this fashion, may turn out to be realistic. 

An exciting implication of this scenario is that RGs inject a substantial
amount of energy in the form of magnetic fields into the cosmic web of filamentary
IGM at $z \sim 2$.
Equipartition field strengths of $10^{-8}$G can arise (GKW), and are in accord with
recent rotation measure based estimates for fields within the denser parts
of the IGM (Ryu et al.\ 1988).
It is well worth noting that two very recent independently advanced arguments, 
starting from essentially
orthogonal evidence based on overall energetics (Kronberg et al.\ 2001;
 Furlanetto \& Loeb 2001)
came to very similar conclusions about the importance of RGs for the injection
of magnetic energy and for galaxy and structure formation.  It is clear
that our ongoing investigations of  aspects of this scenario will
be worthwhile.

\acknowledgements
PJW is grateful for support from RPE funds at GSU.

\end{document}